\documentclass[10pt,preprint2]{aastex}
\shorttitle{Hot subdwarfs and white dwarfs in M4}
\shortauthors{Mochejska et al.}
\begin{document}

\title{Clusters AgeS Experiment. Hot subdwarfs and luminous white
dwarf candidates in the field of the globular cluster
M4\footnote{Based on observations collected at the Las Campanas
Observatory 2.5m du Pont telescope.}}

\author{B.~J. Mochejska, J. Kaluzny}
\affil{Copernicus Astronomical Center, Bartycka 18, 00-716 Warszawa}
\email{mochejsk@camk.edu.pl, jka@camk.edu.pl}
\author{I. Thompson}
\affil{Carnegie Observatories, 813 Santa Barbara Street, Pasadena, 
CA 91101, USA} 
\email{ian@ociw.edu}
\author{W. Pych}
\affil{Copernicus Astronomical Center, Bartycka 18, 00-716 Warszawa}
\email{pych@camk.edu.pl}

\begin{abstract}
We present $UBV$ color magnitude diagrams (CMDs) for the globular
cluster M4. The CMDs show a sequence of four luminous blue stars ($V<20$,
$U-V<-0.6$) which appear to be cluster hot subdwarfs. We present
spectra for the three brightest ones. We also note the presence of a
population of faint blue objects, likely to be hot, young white dwarfs
(WDs) belonging to the cluster. We have selected five objects above
$V=22$ mag, bright enough for follow-up ground-based spectroscopy and
present their coordinates and finding charts. We show a spectrum for
variable V46 (Kaluzny et al. 1997) which suggests that it is a hot
subdwarf, along with a new light curve obtained with the ISIS image
subtraction package (Alard 2000). The light curve is unstable, but
only one period of variability is apparent. Two new variables have
been discovered, both located on the cluster red giant branch
(RGB). We also present a differential $E(B-V)$ reddening map and a
fiducial sequence for the main sequence, subgiant branch and red giant
branch on the $V/B-V$ CMD for a selected region with uniform
reddening. Based on a comparison with the M5 fiducial sequence we
obtain a reddening estimate of $E(B-V)=0.41$ mag towards M4,
consistent with previous determinations.
\end{abstract}
\keywords{globular clusters: individual (M4) -- Hertzsprung-Russell
diagram -- stars: subdwarfs -- stars: white dwarfs }

\section{Introduction}
The Clusters AgeS Experiment (CASE) is a long term project with the
main goal of determining accurate ages and distances to globular
clusters using detached eclipsing binaries (Paczy\'nski 1997). The
photometry of new eclipsing binaries in M4 has already been presented
by Kaluzny et al. (1997). In this paper we investigate the presence of
blue stars in the field of the M4 globular cluster: hot blue subdwarfs
and young bright white dwarfs (WD).

Recently several metal-rich globular clusters have been found to
possess a handful of blue subdwarfs (Rich et al. 1997). Several
scenarios have been proposed to explain the formation of such objects
(Moehler 2001). The study of the properties of these stars could shed
some light on the mechanisms of their creation. Only one such star,
Y453 (Cudworth \& Rees 1990), has been reported to date in M4.

The globular cluster M4 = NGC 6121 ($\alpha_{2000}=16^h23^m36^s,
\delta_{2000}=-26^{\circ} 31\arcmin 32\arcsec$) is the nearest
globular cluster. Due to its proximity M4 has been the subject of
searches for WDs in the past. Several WD candidates were found by Chan
and Richer (1986) in a blink survey designed to select
ultraviolet-excess objects. Drukier, Fahlman \& Richer (1989)
presented the spectra of two of those objects.  One was found to be a
very hot subdwarf (sdO) and the other a field DA with weak hydrogen
lines. Recently a large sample of WDs has been identified with the HST
(Richer et al. 1995; 1997).

Our $UBV$ photometry, although inferior in depth to that obtained with
the HST, covers a much larger area of M4 and thus offers a possibility
to uncover new young bright WD candidates at the top of the cooling
sequence, located in less crowded regions. Such objects would be
within the reach of ground-based spectroscopy. The study of these
objects could provide further constraints on stellar evolution
models of WD progenitors. Our photometry also covers the entire
central part of the cluster, not investigated in the Chan \& Richer
blink survey.

This paper is organized as follows: in Section 2 we describe the
observations. Section 3 outlines the data reduction procedure. In
Section 4 we provide details of the calibration of our instrumental
magnitudes to the standard system. In Section 5 we present and examine
the color-magnitude diagrams (CMDs) constructed from our $UBV$
photometry. In Subsections 5.1 and 5.2 we describe the luminous hot
subdwarfs and the WD candidates, respectively. A differential
reddening map and a fiducial sequence for the $V/B-V$ CMD are
presented in Subsections 5.3 and 5.4. In Subsection 5.5 we determine the
value of the reddening $E(B-V)$ towards M4. We discuss the variable
V46 in Section 6 and present two new variables in Section 7. Section 8
contains the concluding remarks.

\section{Observations}
The photometric data were obtained with the 2.5 m du Pont telescope at
Las Campanas Observatory, equipped with a thinned $2048\times2048$
Tektronix CCD with a scale of $0\farcs 26$ pixel$^{-1}$. An $8\farcm
8\times8\farcm 8$ field centered on the cluster was monitored during
five consecutive nights, from June 1st to 5th, 1995. Observations were
made through Johnson $UBV$ filters. Most of the frames were collected
in the $B$ band, with exposure times varying slightly with the
observing conditions, typically 120s in $B$ and 60s in $V$. The
average seeing was $1\farcs 1$. Newly
discovered variables were presented in Kaluzny et al. (1997). A more
detailed description of the collected data is presented therein.

Spectra of the blue subdwarfs B1, B2 and B3
were obtained at the Baade 6.5 m telescope with the Boller \& Chivens
spectrograph equipped with Tek\#1 CCD. The wavelength coverage extends
from 3774 to 4492 \AA at a resolution of 0.7 \AA. Two spectra were taken for each of the
stars, on May 8th, 2001 for B1 (500 s) and B2 (600 s) and on June 16th
and 19th, 2001 for B3 (1200 s). The spectra were combined to increase
the signal to noise ratio (S/N) which ranged from $\sim16$ to
$\sim 23$ for single exposures.

A 30-minute 6.3 \AA~resolution spectrum of V46, extending from 3753
to 6961 \AA~was obtained on the night of April 25th, 1996 with the
3.6m ESO telescope using the EFOSC1 spectrograph.

\section{Data reduction}

The photometry was carried out using Daophot/ Allstar (Stetson
1994). The positional variability of the point-spread function was
modeled with a second order approximation. The 
extraction of profile photometry closely followed the procedure
described in Mochejska \& Kaluzny (1999).

\begin{figure}[ht]
\plotone{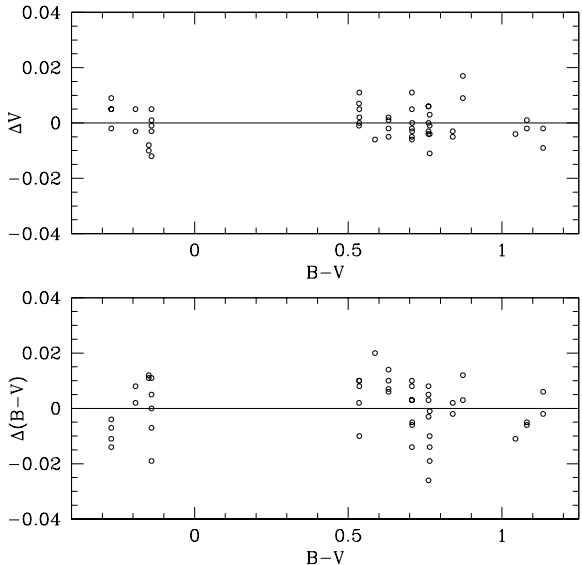}
\caption{$V$ and $B-V$ residuals of Landolt standard stars 
as a function of $B-V$ resulting from transformations defined by
Eqs. (\ref{eq:v98})-(\ref{eq:bv98}). }
\label{fig:res98}
\end{figure}

The $V$ band photometry was obtained from an image combined from three
300 s exposures with subarcsecond seeing. The images were
transformed to a common coordinate system using the ISIS image
subtraction package (Alard 2000) and median combined with
IRAF\footnote{IRAF is distributed by the National Optical Astronomy
Observatories, which are operated by the Association of Universities
for Research in Astronomy, Inc., under cooperative agreement with the
NSF.} using an averaged sigma clipping algorithm to reject cosmic
rays. To recover some of the saturated stars, the photometry from an
averaged $3\times60$ s image, a 20 s and a 6 s exposure was included.  A
similar approach was followed for the $U$ and $B$ band data. The
final $U$ band photometry was derived from a median combined
$4\times900$ s image and a 480 s exposure. For the $B$ band we used an
averaged $2\times800$ s image, a median combined $3\times120$ s image
and a 30 s exposure. The full list of exposures used in the
construction of the CMD is provided in Tab. \ref{tab:exp}.

\begin{figure}[ht]
\plotone{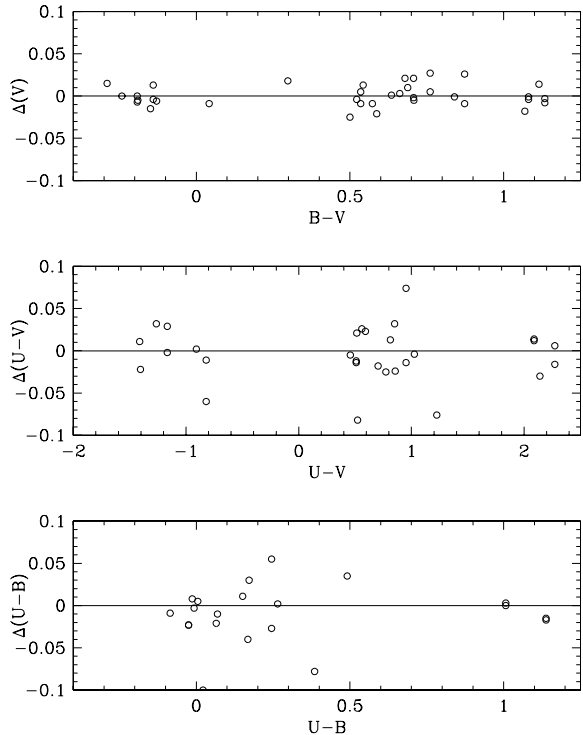}
\caption{$V$, $U-B$ and $U-V$ residuals of Landolt
standard stars as a function of color resulting from transformations
defined by Eqs. (\ref{eq:v40})-(\ref{eq:uv40}).}
\label{fig:res96}
\end{figure}

The photometry lists derived from the images for each filter were
transformed to a common instrumental system defined by the longest
exposure in each filter. Saturated stars were removed from the long
exposures. From the shorter exposure lists we removed stars fainter
than the magnitude at which the average formal error on the short
exposure was twice the error on the longest exposure. In the final
step the lists were added, with the best of the available measurements
(i.e., having the smallest formal error) retained for stars present on
more than one list. In addition we removed 24 RR Lyrae variables from
the photometry lists.

\section{Calibration}

The color terms of the transformations of instrumental magnitudes to
the standard system were derived from 19 observations of Landolt
(1992) fields collected on 5 nights spread between April 22/23 and
June 3/4, 1995. The same instrumental setup (CCD+filters) was used on
all nights. The following values were obtained by averaging color
terms derived for separate nights:
\begin{eqnarray}
\label{eq:v95}
v=V-0.0160(B-V)+c_1\\ 
\label{eq:bv95}
b-v=0.9440(B-V)+c_2\\  
u-b=0.9480(U-B)+c_3\\
u-v=0.9450(U-V)+c_4
\end{eqnarray}

\begin{figure}[htp]
\plotone{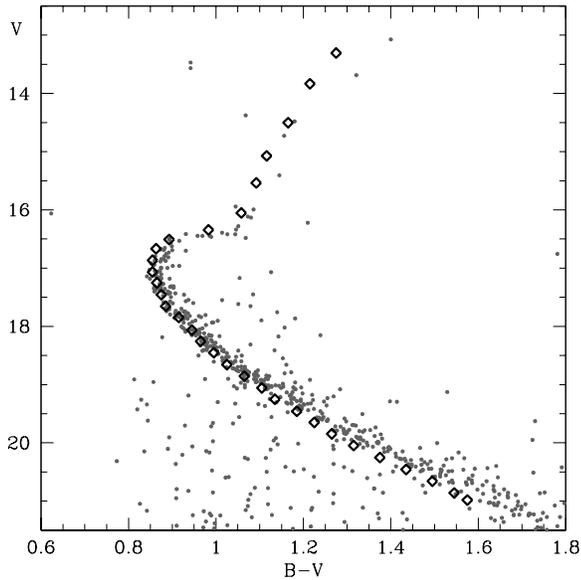}
\caption{A comparison of our $BV$ photometry with the photometry
of Alcaino et al. (1997). Fiducial points determined from our data
(diamonds) are overlaid on the Alcaino et al. CMD (points).}
\label{fig:samus_cmp}
\end{figure}

The constant offsets for $V$ and $B-V$ were determined relative to M4
data collected with the same telescope on the night of May 31/June 1
1998 at airmasses of 1.07-1.11. The transformation coefficients
were derived from 56 stars in 6 Landolt fields observed that night
at airmasses ranging from 1.07 to 1.75. Figure \ref{fig:res98}
shows the residuals of the standard stars resulting from the following
transformations:
\begin{eqnarray}
\label{eq:v98}
v & = & V + 0.6903 - 0.0176 \cdot(B-V)\\
  & &   + 0.1284 \cdot(X-1.25)\nonumber \\
\label{eq:bv98}
b-v &=&    0.3057 + 0.9507 \cdot(B-V) \\
    & & + 0.0731 \cdot(X-1.25)\nonumber 
\end{eqnarray}

\begin{figure}[htp]
\plotone{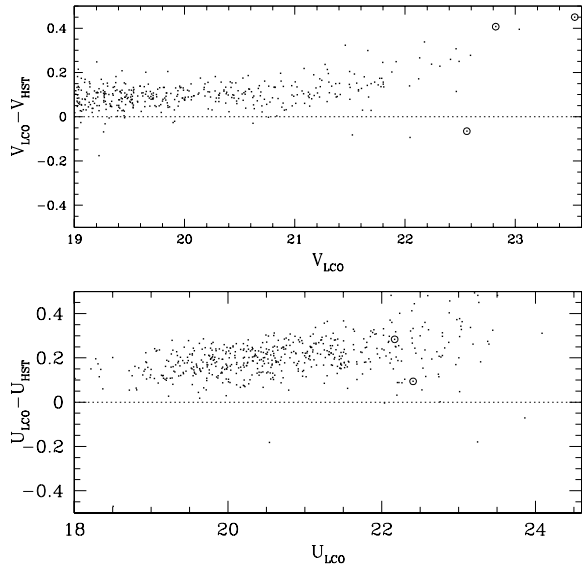}
\caption{A comparison of our $UV$ photometry with the HST photometry
of Ibata et al. (1999). WDs are marked with open circles.}
\label{fig:hst_cmp}
\end{figure}

The offsets for $U-B$ and $U-V$ were determined from a comparison with
M4 data collected with the 1m Swope telescope at LCO at  airmasses
less than 1.06. We observed 39 stars in 9 Landolt fields at airmasses
ranging from 1.07 to 1.36. The extinction coefficients were assumed as
the average values for the winter season at LCO. The following 
transformations were adopted:
\begin{eqnarray}
\label{eq:v40}
v=V+2.7837-0.0189(B-V)+0.15(X-1.25)\\
\label{eq:bv40}
b-v=0.3058+0.9359(B-V)+0.12(X-1.25)\\
\label{eq:ub40}
u-b=2.0252+0.9377(U-B)+0.31(X-1.25)\\
\label{eq:uv40}
u-v=2.3269+0.9313(U-V)+0.44(X-1.25)
\end{eqnarray}
Figure \ref{fig:res96} shows the $V$, $U-B$ and $U-V$ transformation
residuals of the standard stars. 

As a check of consistency of the $V$ and $B-V$ transformations we have
compared the May 31/June 1 1998 calibrated M4 data
(Eqs. (\ref{eq:v98}) and (\ref{eq:bv98})) with the June 1995 data
transformed to the standard system using the color terms in
Eqs. (\ref{eq:v95}) and (\ref{eq:bv95}) and offsets relative to the
transformations given by Eqs. (\ref{eq:v40}) and (\ref{eq:bv40}). For
stars with $V < 16$ mag the offsets in $V$ and $B-V$ are -0.026 and
-0.020 mag, respectively. There is no correlation between the
residuals and color.

We have compared our $V/B-V$ diagram with a CMD constructed from the
photometry obtained by Alcaino et al. (1997) in a $6\arcmin \times
6\arcmin$ field bordering our field on the west (data courtesy of Dr
N. Samus). As Alcaino et al. report the existence of a reddening
gradient across the face of the cluster, with the reddening increasing
to the west, we selected for the comparison the westernmost
$2\farcm 2 \times 8\farcm 8$ and easternmost $2\arcmin \times
6\arcmin$ sections of our and Alcaino et al. fields, respectively. 
We fitted fiducial points to the two datasets in the magnitude range
$16.8 < V < 20$. The mean offset between the two sequences is -0.007
mag. In Figure \ref{fig:samus_cmp} we overplot the Alcaino et al. $V/B-V$
CMD with fiducial points fitted to our data.

In Figure \ref{fig:hst_cmp} we compare our $UV$ photometry with that
obtained with the HST (Ibata et al. 1999; Richer et al. 1997). The
offsets in the photometry are 0.095 mag in $V$ and 0.129 in $U$ for
stars with $19<V<21$ mag and $18<U<22$ mag, respectively. A similar
systematic offset between ground-based and HST $V$ magnitudes has
been noted by Richer et al. (1997; Figure 2 therein). For stars above
$V=20$ mag the difference in photometry dramatically increases with
magnitude. 

\section{Color-magnitude diagrams}

\begin{figure*}
\epsscale{2}
\plotone{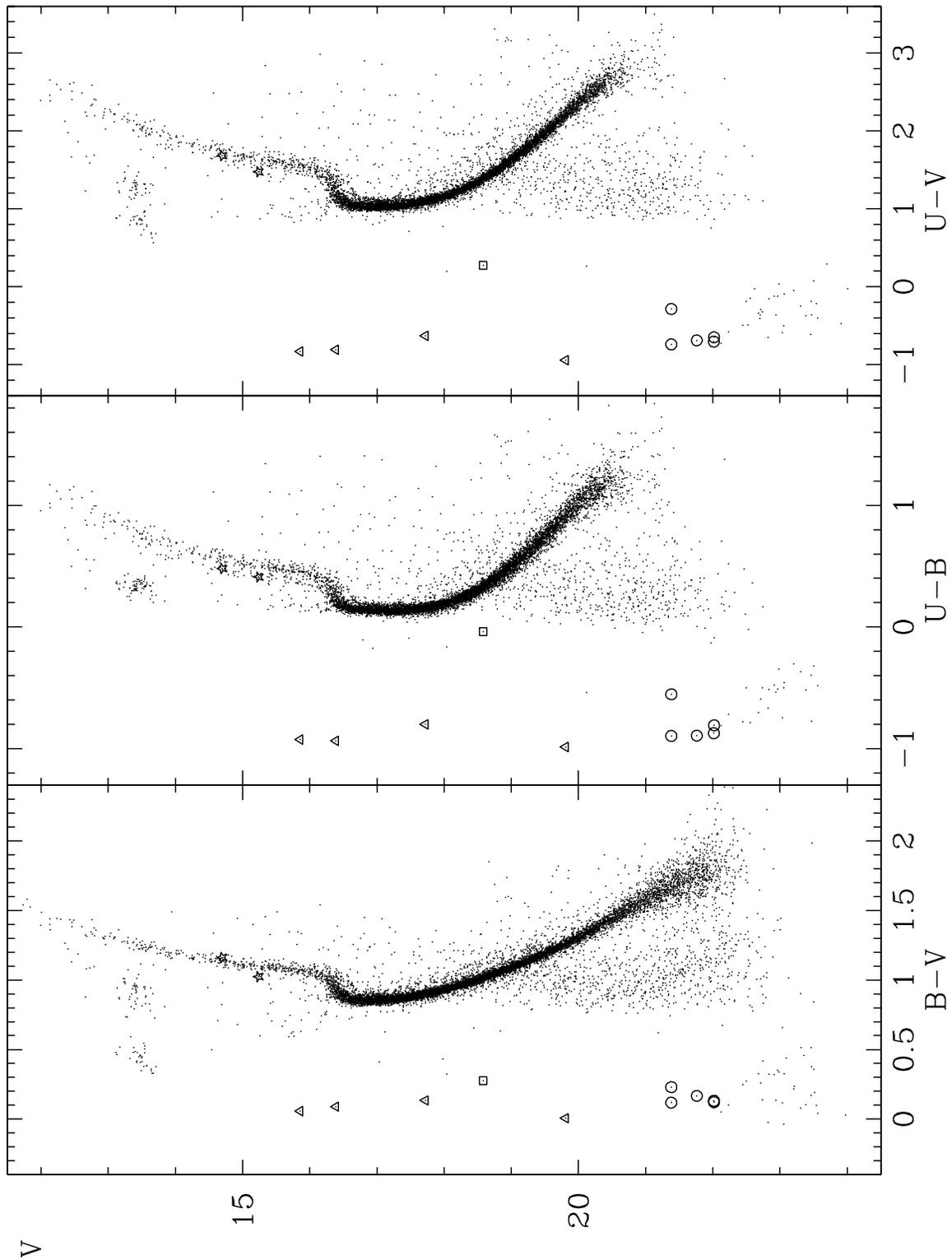}
\caption[]{The $V/B-V$, $V/U-B$ and $V/U-V$ color-magnitude diagrams
of the M4 globular cluster.}
\label{fig:cmd}
\epsscale{1}
\end{figure*}

The $V/B-V$, $V/U-B$ and $V/U-V$ color-magnitude diagrams (CMD)
constructed from our data are presented in Figure \ref{fig:cmd}.  The
CMDs exhibit a well defined main sequence that can be traced down to
$V\simeq22$ mag on the $V/B-V$ CMD and $V\simeq21$ mag on the remaining
two diagrams. The subgiant, red giant and horizontal branches are  well
populated, but there are few stars on the asymptotic giant branch. The
horizontal branch exhibits a clear RR Lyrae gap. A sequence of four
luminous blue stars is present on the CMDs (plotted as open triangles).
These objects are discussed in more detail in Subsection
\ref{sub:blue}. There is also a population of faint blue stars with
magnitudes and colors ($V<21$ mag, $U-V<0.4$ mag) consistent with the
values expected for the upper part of the white dwarf cooling sequence.
A brief discussion of these objects is presented in Subsection
\ref{sub:wd}. In Subsections \ref{sub:red} and \ref{sub:fid} we present
a differential reddening map and a sequence of fiducial points
determined for the $V/B-V$ CMD. In Subsection \ref{sub:ebv} we estimate
the reddening $E(B-V)$ towards M4.

\subsection{Hot subdwarfs}
\label{sub:blue}

We note a sequence of four luminous blue stars on the CMDs ($V<20$,
$U-V<-0.6$), most probably hot subdwarfs belonging to the cluster
(marked with triangles on the CMDs in Figure \ref{fig:cmd}). The
brightest of these, B1, was initially discovered by Cudworth \& Rees
(1990; Y453 in their catalog). No other stars as blue as these were
found in a blink survey aimed at the detection of UV excess objects in
an area between $3\farcm9$ and $23\farcm7$ from the cluster center
(Chan \& Richer 1986). Four bright blue objects with $V<20$ and
$U-V<-0.1$ were found in that study: 1710 (reported as non-stellar),
774, 1285 and 710 (see Table 2 therein). The star 774 was investigated
spectroscopically by Drukier et al. (1989) and found to be most likely
a field WD.  One fainter object, 831 with $V=20.33$ and $U-V=-0.97$
was found to be a very hot subdwarf by Drukier et al. (1989).

\begin{figure*}[ht]
\epsscale{2}
\plotone{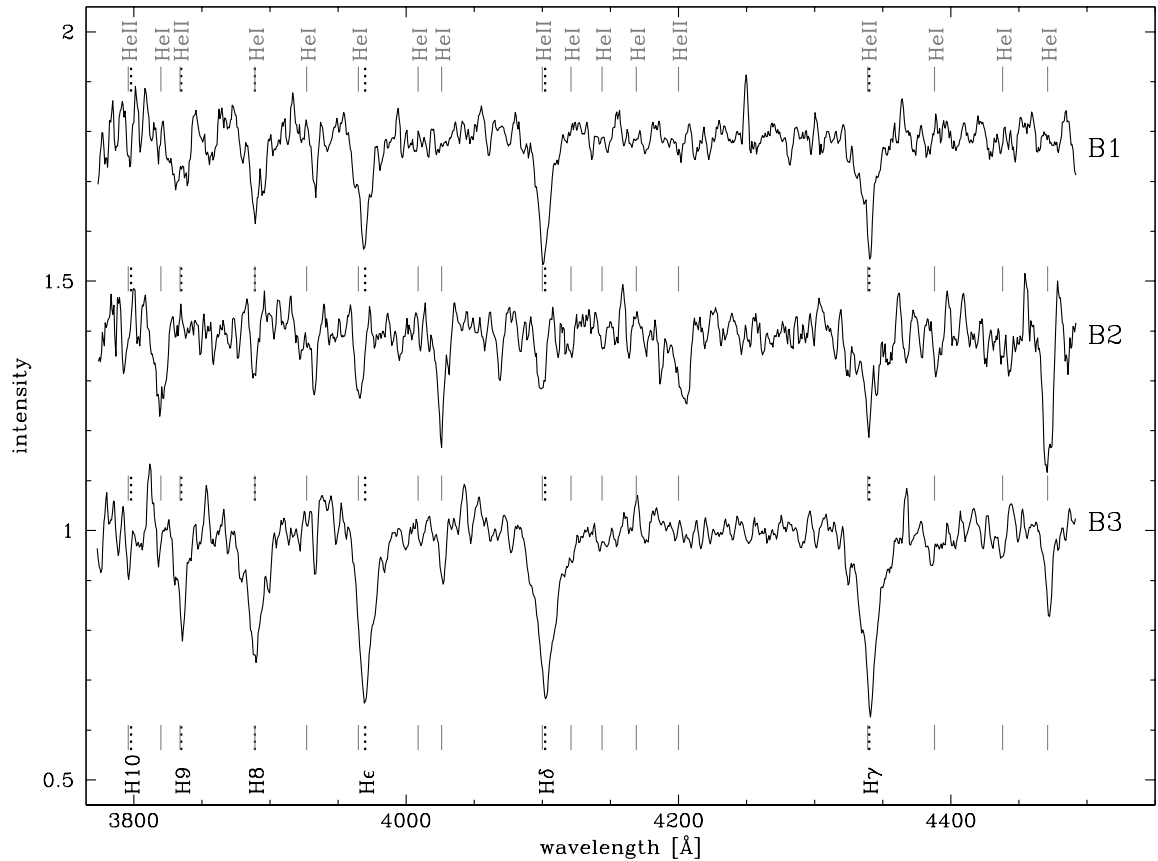}
\caption{The spectra of blue stars B1, B2 and B3, smoothed with a
boxcar filter 5 units in length. He I and He II lines are indicated
with continuous lines, H I with dotted ones.}
\label{bspec}
\epsscale{1}
\end{figure*}

In Figure \ref{bspec} we present the spectra of B1, B2 and B3,
smoothed with a boxcar filter 5 units in length. Helium lines are
indicated with continuous lines and hydrogen lines with dotted
ones. B1 (Y453) has been previously studied spectroscopically by
Moehler, Landsman \& Napiwotzki (1998). Their analysis indicates that
it could be a post-early asymptotic giant branch (PEAGB) star with
$T_{eff}=58800$ K. Our spectrum shows broad Balmer lines, but the He I
and He II lines reported in the Moehler et al.~(1998) data are barely
discernible, most likely due to our higher resolution and moderate
S/N. B2 shows singly and doubly ionized helium lines. The lines at
wavelengths corresponding to the hydrogen Balmer series are most
likely due to doubly ionized helium, judging by the presence of the
\ion{He}{2} line at 4200 \AA. It appears to be a very hot helium rich
subdwarf, approximately sdO5:He4 in the Jeffery et al. (1997)
classification scheme. The precise classification is somewhat
uncertain, as it is based on only three lines in the 300 \AA~overlap
between our and their spectra: the \ion{He}{2} line at 4339 \AA~and
\ion{He}{1} lines at 4388 \AA~and 4471 \AA. The spectrum of B3
exhibits broad Balmer lines and two lines from singly ionized helium
(4026 \AA~and 4471 \AA) and none from \ion{He}{2}. It appears to be an
sdB:He1 subdwarf.

The spectrum of B2, which appears to he helium-rich, is quite unusual.
Hot subdwarfs in globular clusters are almost always found to be
helium-poor (e.g.~Moehler et al.~1997a). Only one hot helium-rich
subdwarf in a globular cluster has been reported to date (Moehler et
al.~1997b). B2 seems to be only the second such star known. Brown et
al.~(2001) suggest that such stars might have undergone late
helium-core flash while descending the WD cooling sequence. A
convection zone produced during the helium flash would mix hydrogen
into the hot interior, where it would be consumed, thus greatly
enhancing the envelope helium abundance. However, B2 seems to be too
luminous for their scenario. 

The hot subdwarf B3 and possibly B2 could belong to a vertical
extended horizontal branch (EHB), not reported in previous studies of
this cluster. Generally the globular cluster horizontal branch (HB)
morphology correlates with metallicity - the HB becomes redder as the
mean metal abundance increases. However, variations in HB morphology
at a given metallicity are observed and a second, still unknown
parameter has to be introduced. It has been suggested that mass loss
on the red giant branch may be the second parameter (Catelan
2000). The stars which suffered strong mass loss would become EHB
stars (D'Cruz et al. 1996, D'Cruz 1998). Recently the existence of
high-metallicity hot blue horizontal branch stars has been reported in
the relatively metal-rich globular clusters NGC 1851 (Bellazzini et
al.~2001) and 47 Tuc (Ferraro et al.~2001), the very old metal-rich
open cluster NGC 6791 (Kaluzny \& Udalski 1992) and the Milky Way
bulge (Peterson et al. 2000). Detailed study of the nature of such
extreme horizontal branch stars could provide valuable clues regarding
the unknown second parameter.

An alternative explanation for B4 is that it is a low-mass helium
WD. Similar stars have been identified in the globular cluster NGC
6397 (Cool et al. 1998) and the spectrum of one of them was found to
be consistent with a He WD (Edmonds et al. 1999). Using 
differential surface density data from HEQS (K\"{o}hler et al. 1997),
LBQS (Hewett et al. 1995) and the 2dF QSO survey (Boyle et al. 2000) we 
estimate the probability of B4 being a quasar to be between 2\%
and 48\%.

We have checked the subdwarf candidates for variability in our
observations spanning 5 days and found them to be constant within the
errors of the photometry.

\begin{figure}[ht]
\plotone{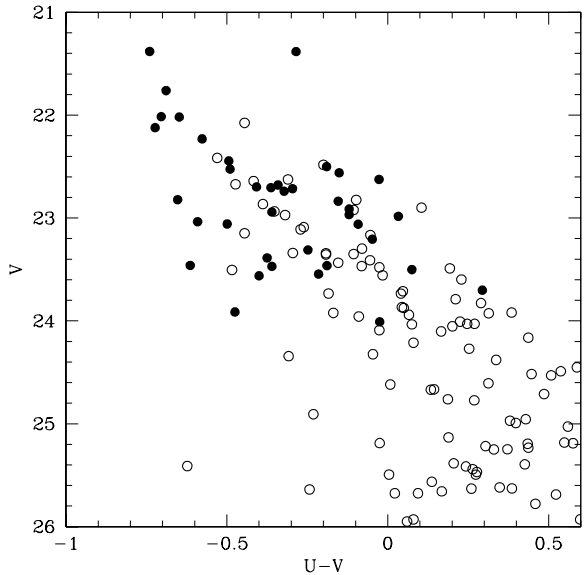}
\caption{A $V/U-V$ CMD showing our white dwarfs (filled circles) 
and those discovered with the HST by Richer et al. (1997, open circles).}
\label{fig:wd_cmp}
\end{figure}

In Table \ref{tab:blue} we list the photometry and equatorial
coordinates for the hot subdwarfs. 
Finding charts for these stars are are presented in Figure \ref{fig:blue}.
Each chart covers $33\arcsec \times 33\arcsec$ on the sky.

\subsection{White dwarf candidates}
\label{sub:wd}

A population of faint blue stars is present on our CMDs. In
Figure \ref{fig:wd_cmp} we overplot these stars (filled circles) with
the WDs observed with the HST (open circles, Richer et
al. 1997). Their location is consistent with the upper part of the
HST white dwarf cooling sequence. Figure \ref{fig:hst_cmp} is
a comparison of our $UV$ photometry and that of the HST, 
open circles indicate WDs that are common to the two studies.

\begin{figure}[ht]
\plotone{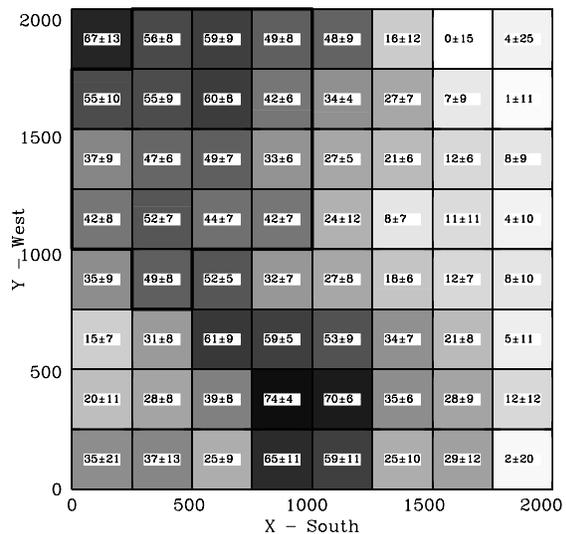}
\caption{The differential reddening map for 64
$1\farcm1\times1\farcm1$ subfields covering the field of M4.  The
differential reddening is given in millimagnitudes measured relative
to the section with the lowest reddening. Higher values of $\delta
E(B-V)$ correspond to higher reddening. The region used in the
determination of the fiducial sequence is shown surrounded with a
thick line. The cluster center is located approximately at
$(X_C,Y_C)=(1100,1000)$. }
\label{fig:ebv}
\end{figure}

The photometry and equatorial coordinates for the five brightest
objects with $V \la 22$ (marked with open circles in Figure
\ref{fig:cmd}) are presented in Table
\ref{tab:wd}\footnote{The full table is available from the authors via
{\ttfamily anonymous ftp} from {\ttfamily cfa-ftp.harvard.edu} in the
directory {\ttfamily /pub/bmochejs/M4/} } and $33\arcsec \times
33\arcsec$ $V$ and $U$ band finding charts are presented in Figure
\ref{fig:map_wd}. These objects are located in the outer, less crowded
regions of the cluster (compared to the HST WD sample) and are bright
enough for ground-based spectroscopy.

\begin{figure}[ht]
\plotone{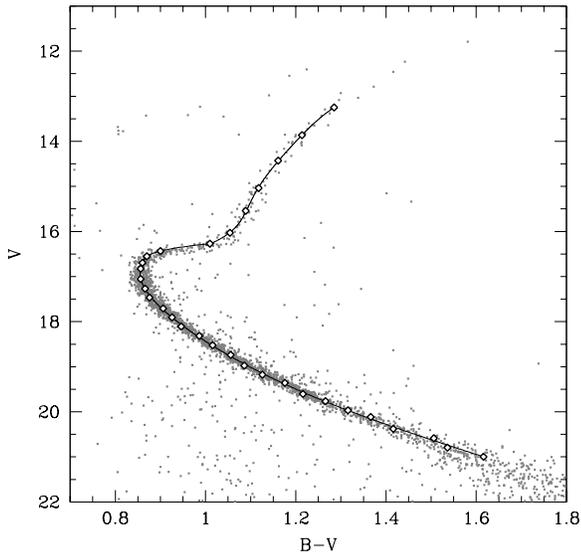}
\caption{The fiducial sequence for the $V/B-V$ CMD in the
selected region of M4 with uniform absorption.}
\label{fig:fid}
\end{figure}

\subsection{Differential Extinction Map}
\label{sub:red}

All of the principal sequences on the CMDs display a broadening due to
the differential reddening across the face of the cluster, as
discussed by Cudworth \& Rees (1990), Fahlman et al. (1996), and
Alcaino et al.~(1997). The differential reddening has been
investigated by Lyons et al.~(1995) using K{\sc i} column density
measurements towards 16 stars and by Ivans et al. (1999; 2000) on the
basis of spectroscopic analysis of 22 red giant branch stars. These
studies revealed the existence of a patchy reddening variation across
the cluster face. In addition, Peterson, Rees \& Cudworth (1995) found
evidence for an unusually high value of $R_V=A_V/E(B-V)\approx4$.

We have constructed an $8\times8$ element differential reddening map
following closely the procedure presented by von Braun \& Mateo
(2000), based on the approach of Kaluzny \& Krzeminski (1993). The
errors were also derived as described therein, taking into account the
correlation between the errors in color and magnitude. We adopted
$R_V=4$ in our analysis. Figure \ref{fig:ebv} presents the derived map
of differential reddening, measured relative to the section with the
lowest reddening.  The extinction appears to be weakest in the east
and increases to the west, in accordance with results obtained in
previous studies (eg. Alcaino et al. 1997). The patchy nature of the
reddening is also confirmed.

\begin{figure}[htp]
\plotone{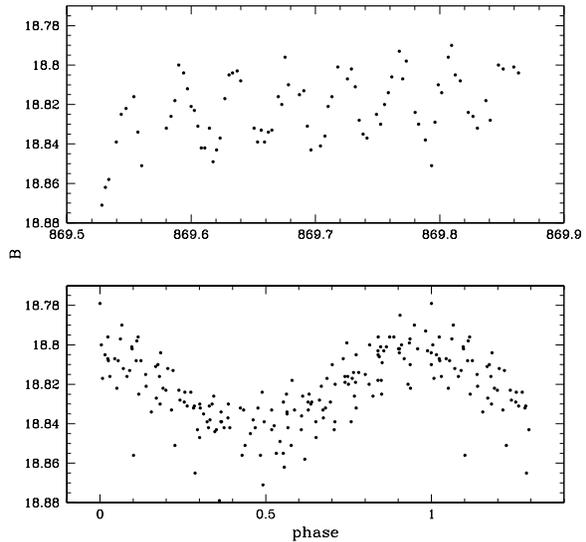}
\caption{The light curve of the hot subdwarf V46. Upper panel: the raw 
light curve for the first night of observations. Lower panel: light curve
from all five nights phased with a period of 0.04358 d}
\label{fig:lc_46}
\end{figure}

We have compared our differential extinction map with the Schlegel,
Finkbeiner \& Davis (1998; hereafter SFD) dust map. There is some
qualitative similarity between the two maps: both show lowest
reddening in the northeastern corner, intermediate in the southwestern
and higher in the northwestern corners. The maps disagree as to the reddening
in the southeastern section: the SFD map shows the strongest reddening
there, while it is rather small on our map. Another difference is that
the differential reddening is larger on our map (up to 0.074) than on
the SFD map (up to 0.039). We have checked a larger area around M4 on
the SFD map and noted that it is located in a region of highly
variable extinction. The cluster coincides with the center of a void
with patchy lower extinction ($E(B-V)\sim 0.5$) some $1^{\circ} \times 
1^{\circ}$ surrounded by thick filaments of higher extinction ($E(B-V)>1$
mag). Moreover, we notice that the $E(B-V)$ values from the SFD map
range from 0.472 to 0.511 mag for the cluster area, which is
substantially higher than the reddening value of $E(B-V)=0.36$ mag
reported for M4 by Harris (1996). This discrepancy might be due to the
fact that the SFD map also takes into account the extinction caused by
dust located behind the cluster. This interpretation would invalidate
to large extent any comparison between the two reddening maps.

\subsection{$V/B-V$ CMD Fiducial Sequence}
\label{sub:fid}

To derive the $V/B-V$ fiducial sequence we have selected a region of
uniform reddening using the differential reddening map constructed in
Subsection \ref{sub:red}. This region is shown in Figure \ref{fig:ebv}
surrounded with a thick line. The fiducial points were determined
separately for the main sequence (MS; $16.75 < V < 21.0$), the
subgiant branch (SGB; $16.2 < V < 16.75$) and the red giant branch
(RGB; $13.0 < V < 16.2$). For the MS and SGB, the fiducial points were
determined by finding the mode of the color distribution, smoothed
with a boxcar filter 0.03 mag in length, in magnitude bins 0.21 and
0.135 mag wide, respectively. The magnitude for each bin was computed
as the mean. In case of the RGB we adopted a different procedure, as
this branch is more sparsely populated than the previous two. To avoid
contamination of the RGB sample, especially with the AGB and HB stars,
the following magnitude-dependent color cuts were applied:
$(29.7-V)/14.5 < B-V < (32.6-V)/14.5$.  For the RGB, the fiducial
points were computed as the mean color in magnitude bins 0.55 mag
wide. The magnitudes within each bin were determined as the mean, just
as for the MS and SGB. The resultant fiducial sequence is presented in
Figure \ref{fig:fid}, superimposed on the CMD which was used in their
derivation. The fiducial sequence is listed in Table \ref{tab:fid}.

\begin{figure*}[ht]
\epsscale{2}
\plotone{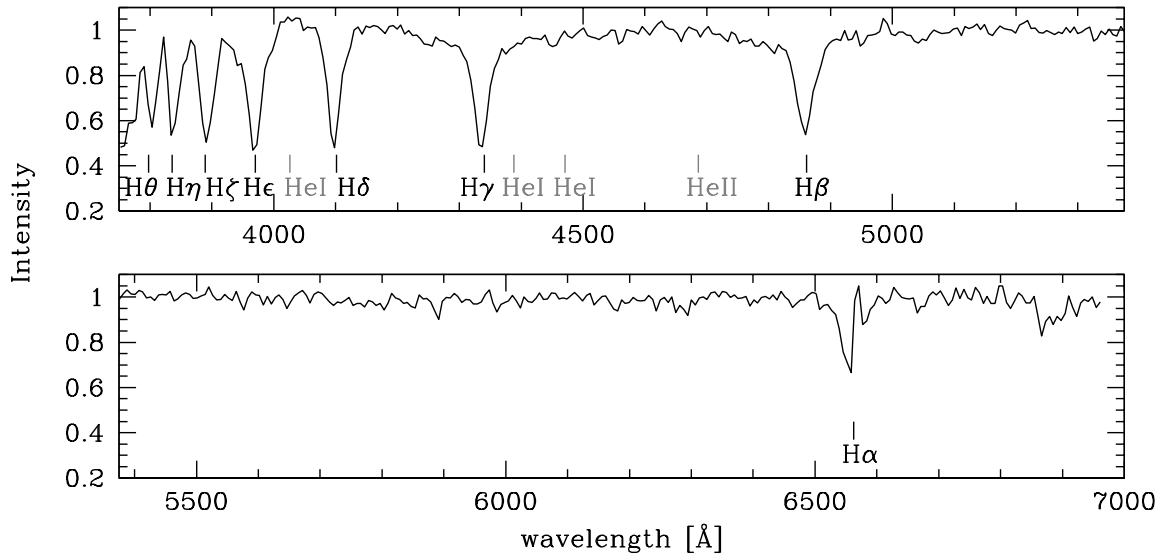}
\caption{The spectrum of the hot subdwarf V46. The hydrogen Balmer
series and helium lines are marked and labeled. The wavelengths of the
helium lines are as follows: \ion{He}{1}(4026\AA)
\ion{He}{1}(4388\AA), \ion{He}{1}(4471\AA), \ion{He}{2}(4686\AA).}
\label{fig:sp}
\epsscale{1}
\end{figure*}

\subsection{Foreground Extinction E(B-V)}
\label{sub:ebv}

We have obtained an independent estimate of foreground extinction in
B-V towards M4 from comparison of the fiducial sequence constructed in
the previous Subsection with the one given by Sandquist et al. (1996)
for the M5 globular cluster. We have chosen M5 for this purpose due to
its low reddening (E(B-V) = 0.03; Harris 1996), similar
metallicity ([Fe/H] = -1.11 dex as compared with -1.19 dex for M4;
Carretta \& Gratton 1997) and similar age ($\tau_{M4} - \tau_{M5} =
0.6$ Gyr; Rosenberg et al. 1999).

To determine the $B-V$ colors of the cluster turnoffs (TO) we fitted
the fiducial points in the turnoff regions with a fifth order Chebyshev
polynomial. The color of the turnoff was taken as the color of the
bluest point on the fitted curve. We obtained the values of 0.854 and
0.470 for M4 and M5, respectively. From the isochrones of Girardi et
al. (2000) we estimated the corrections in $(B-V)_{TO}$ for the
difference in metallicity (-0.08 dex) and age (0.6 Gyr) between M4 and
M5 to be -0.014 and +0.006 mag, respectively. Assuming an
extinction of 0.03 mag towards M5, we obtain a reddening value of
$E(B-V)=0.41$ for M4. This is consistent with the value of 0.36 listed
by Harris (1996), taking into account the variability of the
extinction across the face of the cluster. It is in disagreement with
the total extinction estimate of ~0.5 mag based on the SFD dust maps.

\begin{figure}[ht]
\plotone{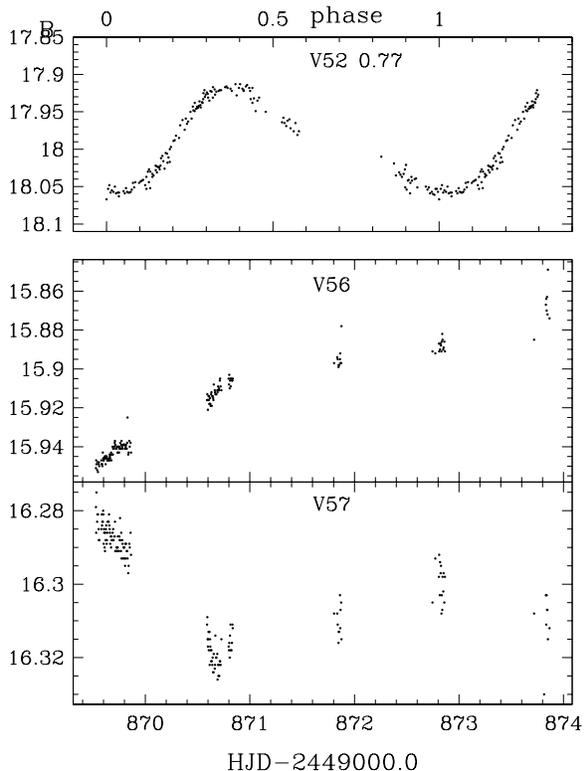}
\caption{The light curves of the variables V52, V56 and V57. The last
two variables are newly discovered. V52 is phased with the newly
derived period of 0.77d.}
\label{fig:lc_3}
\end{figure}

\section{The variable hot subdwarf V46} 
V46 was detected to be variable by Kaluzny et al. (1997). The star
exhibits coherent modulations with a period of 0.04358 d. It is
indicated on the CMDs in Figure \ref{fig:cmd} with an open square and
mean photometry is listed in Table \ref{tab:new}. We have measured the
light curve of V46 using the ISIS image subtraction package (Alard \&
Lupton 1998; Alard 2000). The upper panel of Figure \ref{fig:lc_46}
shows the raw light curve for the first night of observations. The
phased light curve from all five nights is shown in the lower
panel. The shape of the light curve appears to be unstable. We have
searched for other coherent periodicities and have found
none. Nevertheless, we cannot rule out multiple periodicity, due to
the short timespan of our observations.

The spectrum of V46, shown in Figure \ref{fig:sp}, exhibits strong and
broad Balmer lines, while the \ion{He}{1} and \ion{He}{2} lines are
absent. We have compared it with spectra for stars of types B through
F, taken from the spectral atlas of Pickles (1998). Our spectrum is
consistent with types from late B through late A. The dereddened
colors $(B-V)_0=-0.13$ and $(U-B)_0=-0.32$ suggest V46 is a late B
type star. In order to be a main sequence late B star
($M_V\simeq-1.5$), at $V_0\sim17.1$ it would have to be located at a
distance modulus of $\sim18.6$ mag or 52 kpc (assuming the larger
$E(B-V)=0.5$ from Schlegel et al. (1998)). At $b=15.97$ this
corresponds to $Z=14$ kpc. This suggests that V46 is rather a hot
subdwarf, type sdB1 in the Jeffery et al. (1997) classification
scheme. 

Although the spectrum of V46 indicates that it is an sdB, its
period of brightness modulations (0.04358d=3765s) is much longer than
the typical periods of 100-200s (500s in extreme cases) for pulsating
subdwarfs (Koen et al. 1998). 

Another possibility to consider are AM CVn type interacting binary WD
systems. AM CVn itself exhibits a low amplitude flickering light curve
with a period of 0.01216d=1051s (Krzeminski 1972; Provencal et
al. 1995). GP Com (G61-29) is the longest period variable of this
class, with P=0.03231d=2791s (Marsh 1999). However, AM CVn type
variables show He I features in their spectra, accompanied by a lack
of hydrogen lines. This is in disagreement with the V46 spectrum and
thus renders this interpretation very unlikely.

A different interpretation, also suggested more by the light curve
than by the spectrum, is that V46 is a field SX Phe star. Its period
is typical for these variables. The light variations appear remarkably
similar to those of SX Phe stars (i.e. Breger et al.\ 1995, Handler et
al.\ 1996, Pych et al.\ 2001, Kaluzny \& Thompson 2001). The location
of V46 on the CMD (Fig.\ \ref{fig:cmd}) would suggest that it should
be located behind the cluster.

An alternative interpretation could be that V46 is a binary star with
the brightness modulations due to the reflection effect or to
ellipsoidal variations (P=0.08716d=2.09h). A study of the radial
velocities of a sample of 70 sdBs indicates that 45\% of them are
post-common envelope binaries with periods of the order of a few hours
to a few days (Saffer, Green \& Bowers 2000). The period for V46 is
consistent with this period range. Spectra with better time resolution
are needed to verify this interpretation.

\section{Other variables}
\begin{figure*}[hp]
\epsscale{2}
\plotone{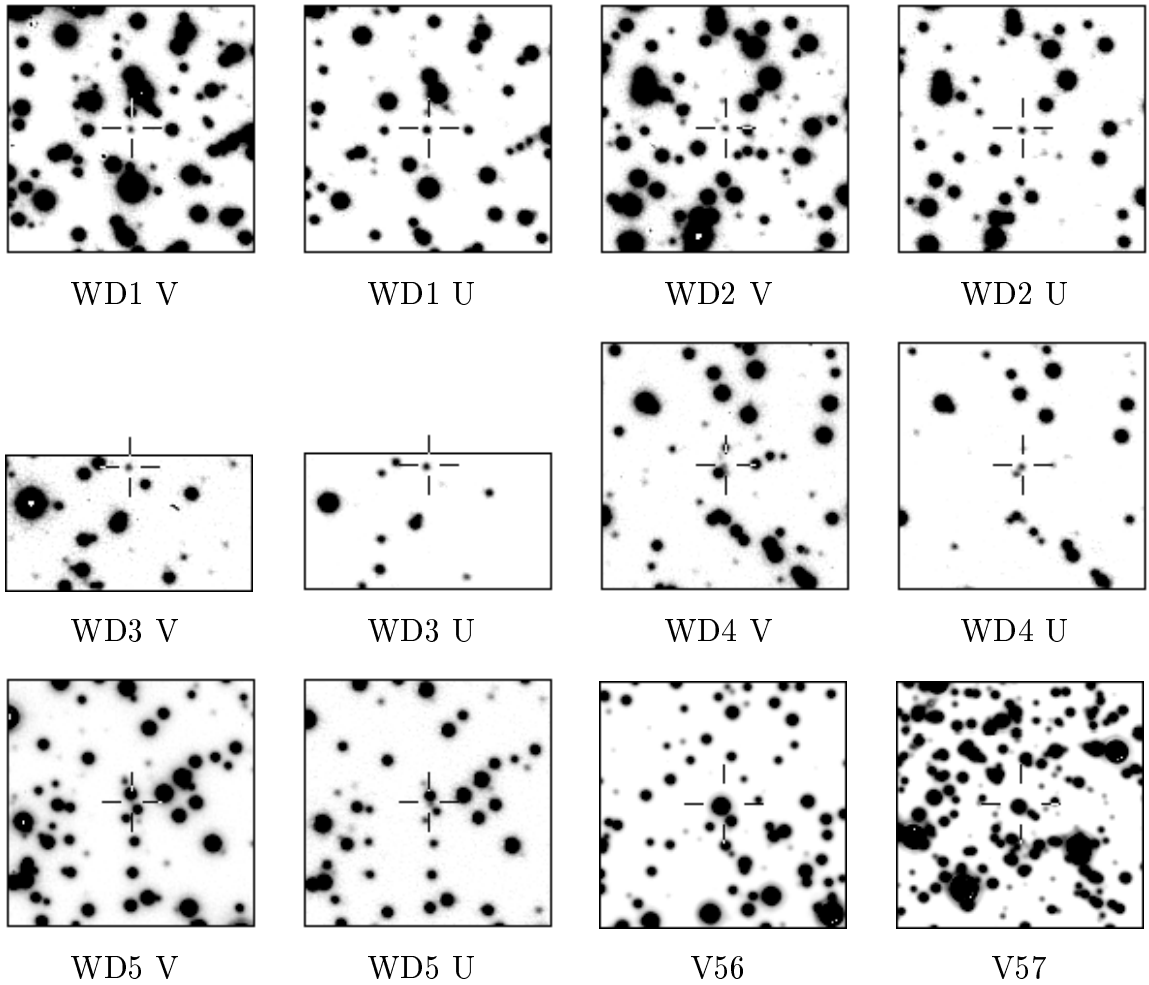}
\caption{Finding charts for the white dwarfs and two new variables,
V56 and V57. North is up and east is to the right.}
\label{fig:map_wd}
\epsscale{1}
\end{figure*}

\begin{figure*}[hp]
\epsscale{2}
\plotone{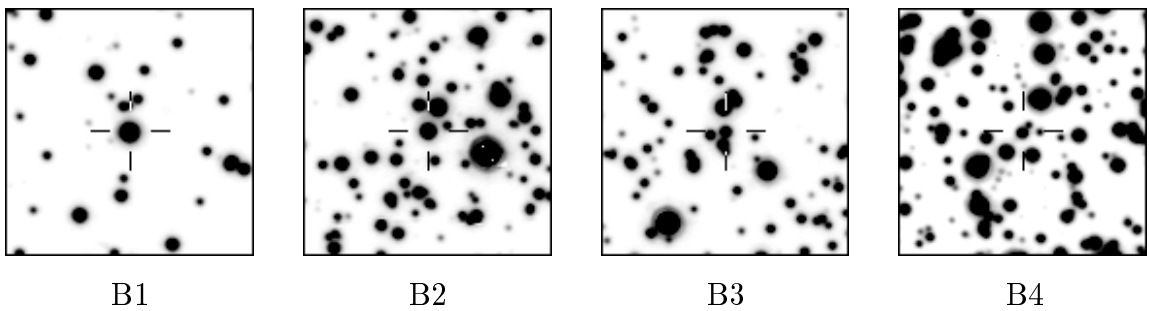}
\caption{Finding charts for the blue subdwarfs. North is up and east is
to the right.}  
\label{fig:blue}
\epsscale{1}
\end{figure*}

The data set was reanalyzed using the ISIS image subtraction package
(Alard \& Lupton 1998, Alard 2000) mainly with the aim of obtaining a
better light curve for V46. The light curves of all previously found
variables were also extracted and two new variables were found. There
is notable improvement in the quality of the photometry extracted
with ISIS, as compared to traditional profile photometry. In
Figure \ref{fig:lc_3} we show the new ISIS light curve for the
previously discovered variable V52, phased with its newly derived
period of 0.7765 days (Tab. \ref{tab:new}). The light curves of the
two new variables V56 and V57 are also presented in
Figure \ref{fig:lc_3}, their photometry is listed in Tab. \ref{tab:new}
and finding charts are presented in Figure \ref{fig:map_wd}. On the CMDs
(Figure \ref{fig:cmd}) these variables are located on the RGB (star
symbols).

\section{Conclusions}

Our investigation of the M4 $UBV$ color magnitude diagrams has
resulted in the identification of four luminous hot subdwarfs (one
previously known) and the tip of the white dwarf cooling sequence.
The hot blue stars could possibly form a vertical extended horizontal
branch of the cluster. The spectra for the three brightest blue stars
confirm that they are hot subdwarfs. Further study would be required
to elucidate the nature of the fourth object. We have selected five
luminous WD candidates above $V\simeq22$ mag, located in the outer,
less crowded regions of M4. These objects would make good targets for
follow-up ground-based spectroscopy. In addition we present a fiducial
sequence for the $V/B-V$ CMD and a differential reddening map. We
obtain an estimate of $E(B-V)=0.41$ for the foreground reddening
towards M4 based on a comparison with M5, a value consistent with
previous determinations.

We also studied the variable star V46, discovered by Kaluzny et
al. (1997) in the same data set. We present and discuss its spectrum,
which indicates that it is a hot subdwarf and present a new light
curve, derived with the ISIS image subtraction package (Alard 2000).
We also identify two new variables, both lying on the cluster RGB. 

\begin{acknowledgements}  
W. Landsman, G. Preston, and K.Z. Stanek provided us with helpful
comments on the manu\-script. We would like to thank H. Duerbeck for
help with the spectroscopic observations, G. Pojma{\'n}ski for $lc$ -
the light curve analysis utility and N. Samus for the comparison $BV$
photometry. We also thank the referee for a very useful report, which
significantly improved this paper. BJM, JK and WP were supported by
the Polish KBN grant 5P03D004.21. BJM was also supported by the
Foundation for Polish Science stipend for young scientists. JK was
also supported by the NSF grant AST-9819787. IT was supported by the
NSF grant AST-9819786.
\end{acknowledgements}

\begin{small}   
\begin{planotable}{crc}
\tablewidth{0pc}
\tablecaption{\sc List of Exposures}
\tablehead{ \colhead{Filter} & \colhead{Exposure Time [s]} & 
\colhead{FWHM [$\arcsec$]}}
\startdata
U & $4\times900$ & 0.91 \\
  & $480$        & 1.17 \\
B & $2\times800$ & 1.01 \\
  & $3\times120$ & 0.91 \\ 
  & $30$         & 0.99 \\
V & $3\times300$ & 0.86 \\
  & $3\times60$  & 1.04 \\
  & $20$         & 1.09 \\
  & $6$          & 0.99 \\
\enddata
\label{tab:exp}
\end{planotable}
\end{small}

\begin{small}   
\begin{planotable}{cccccc}
\tablewidth{0pc}
\tablecaption{\sc Hot subdwarf candidates in M4}
\tablehead{ \colhead{ID} & \colhead{$\alpha_{2000}$} & 
\colhead{$\delta_{2000}$} & \colhead{$V$} & \colhead{$B-V$} & \colhead{$U-V$}}
\startdata
 B1 & 245.84248 & -26.46734 & 15.86 & 0.06 & -0.83 \\
 B2 & 245.85941 & -26.52419 & 16.39 & 0.09 & -0.81 \\
 B3 & 245.86347 & -26.54041 & 17.72 & 0.13 & -0.63 \\
 B4 & 245.89129 & -26.50645 & 19.81 & 0.01 & -0.94 \\
\enddata
\label{tab:blue}
\tablecomments{B1=Y435}
\end{planotable}
\end{small}

\begin{small}
\begin{planotable}{cccccc}
\tablewidth{0pc}
\tablecaption{\sc Luminous white dwarf candidates in M4}
\tablehead{ \colhead{ID} & \colhead{$\alpha_{2000}$} & 
\colhead{$\delta_{2000}$} & \colhead{$V$} & \colhead{$B-V$} & \colhead{$U-V$}}
\startdata
 WD1 & 245.89291 & -26.57736 & 21.38 &  0.12 & -0.74 \\ 
 WD2 & 245.94001 & -26.55283 & 21.38 &  0.23 & -0.29 \\ 
 WD3 & 245.86420 & -26.45166 & 21.76 &  0.17 & -0.69 \\ 
 WD4 & 245.86497 & -26.47151 & 22.02 &  0.12 & -0.65 \\ 
 WD5 & 245.84395 & -26.55067 & 22.02 &  0.13 & -0.70 \\
\enddata
\label{tab:wd}
\end{planotable}
\end{small}

\begin{small}
\begin{planotable}{cc}
\tablewidth{0pc}
\tablecaption{\sc The M4 $V/B-V$ fiducial sequence}
\tablehead{ \colhead{$V$} & \colhead{$B-V$}}
\startdata
 13.250 & 1.285 \\
 13.861 & 1.214 \\
 14.430 & 1.161 \\
 15.034 & 1.118 \\
 15.542 & 1.089 \\
 16.029 & 1.054 \\
 16.270 & 1.010 \\
 16.435 & 0.900 \\
 16.550 & 0.870 \\
 16.695 & 0.860 \\
 16.826 & 0.856 \\
 17.056 & 0.856 \\
 17.266 & 0.866 \\
 17.466 & 0.876 \\
 17.706 & 0.906 \\
 17.906 & 0.926 \\
 18.106 & 0.946 \\
 18.316 & 0.986 \\
 18.526 & 1.016 \\
 18.736 & 1.056 \\
 18.976 & 1.086 \\
 19.176 & 1.126 \\
 19.356 & 1.176 \\
 19.606 & 1.216 \\
 19.766 & 1.266 \\
 19.966 & 1.316 \\
 20.116 & 1.366 \\
 20.376 & 1.416 \\
 20.586 & 1.506 \\
 20.796 & 1.536 \\
 20.996 & 1.616 \\
\enddata
\label{tab:fid}
\end{planotable}
\end{small}

\begin{small}
\begin{planotable}{ccccccc}
\tablewidth{0pc}
\tablecaption{\sc Other variables in M4}
\tablehead{ \colhead{ID} & \colhead{$\alpha_{2000}$} & 
\colhead{$\delta_{2000}$} & \colhead{ P(days)}& \colhead{$V_{max}$} & 
\colhead{$B-V$} & \colhead{$U-V$}} 
\startdata
 46 & 245.94613 & -26.53210 & 0.04358& 18.58 &  0.28 &  0.28 \\
 52 & 245.88068 & -26.51577 & 0.7765 & 16.95 &  0.97 &  1.27 \\
 56 & 245.89239 & -26.49860 & \nodata& 14.70 &  1.16 &  1.68 \\ 
 57 & 245.90251 & -26.53045 & \nodata& 15.24 &  1.03 &  1.48 \\ 
\enddata
\label{tab:new}
\end{planotable}
\end{small}

\end{document}